\documentclass[reprint,amsmath,amssymb,aps,floatfix,superscriptaddress,showpacs,prl,dvipdfmx]{revtex4-1}
\usepackage[dvipdfmx]{graphicx}
\usepackage{graphicx}
\usepackage{dcolumn}
\usepackage{bm}
\usepackage{color}
\usepackage{ulem}
  
\newcommand{\el}{\mbox{${\rm e^{-}}$}}
\newcommand{\ps}{\mbox{${\rm e^{+}}$}}

\begin{document}
\title{Direct Measurement of the Spectral Structure of Cosmic-Ray Electrons+Positrons 
  in the TeV Region with CALET on the International Space Station}
%
\author{O.~Adriani}
\affiliation{Department of Physics, University of Florence, Via Sansone, 1 - 50019, Sesto Fiorentino, Italy}
\affiliation{INFN Sezione di Florence, Via Sansone, 1 - 50019, Sesto Fiorentino, Italy}
\author{Y.~Akaike}
\email[]{yakaike@aoni.waseda.jp}
\affiliation{Waseda Research Institute for Science and Engineering, Waseda University, 17 Kikuicho,  Shinjuku, Tokyo 162-0044, Japan}
\affiliation{JEM Utilization Center, Human Spaceflight Technology Directorate, Japan Aerospace Exploration Agency, 2-1-1 Sengen, Tsukuba, Ibaraki 305-8505, Japan}
\author{K.~Asano}
\affiliation{Institute for Cosmic Ray Research, The University of Tokyo, 5-1-5 Kashiwa-no-Ha, Kashiwa, Chiba 277-8582, Japan}
\author{Y.~Asaoka}
\affiliation{Institute for Cosmic Ray Research, The University of Tokyo, 5-1-5 Kashiwa-no-Ha, Kashiwa, Chiba 277-8582, Japan}
\author{E.~Berti} 
\affiliation{INFN Sezione di Florence, Via Sansone, 1 - 50019, Sesto Fiorentino, Italy}
\affiliation{Institute of Applied Physics (IFAC),  National Research Council (CNR), Via Madonna del Piano, 10, 50019, Sesto Fiorentino, Italy}
\author{G.~Bigongiari}
\affiliation{Department of Physical Sciences, Earth and Environment, University of Siena, via Roma 56, 53100 Siena, Italy}
\affiliation{INFN Sezione di Pisa, Polo Fibonacci, Largo B. Pontecorvo, 3 - 56127 Pisa, Italy}
\author{W.R.~Binns}
\affiliation{Department of Physics and McDonnell Center for the Space Sciences, Washington University, One Brookings Drive, St. Louis, Missouri 63130-4899, USA}
\author{M.~Bongi}
\affiliation{Department of Physics, University of Florence, Via Sansone, 1 - 50019, Sesto Fiorentino, Italy}
\affiliation{INFN Sezione di Florence, Via Sansone, 1 - 50019, Sesto Fiorentino, Italy}
\author{P.~Brogi}
\affiliation{Department of Physical Sciences, Earth and Environment, University of Siena, via Roma 56, 53100 Siena, Italy}
\affiliation{INFN Sezione di Pisa, Polo Fibonacci, Largo B. Pontecorvo, 3 - 56127 Pisa, Italy}
\author{A.~Bruno}
\affiliation{Heliospheric Physics Laboratory, NASA/GSFC, Greenbelt, Maryland 20771, USA}
\author{J.H.~Buckley}
\affiliation{Department of Physics and McDonnell Center for the Space Sciences, Washington University, One Brookings Drive, St. Louis, Missouri 63130-4899, USA}
\author{N.~Cannady}
\email[]{nick.cannady@nasa.gov}
\affiliation{Center for Space Sciences and Technology, University of Maryland, Baltimore County, 1000 Hilltop Circle, Baltimore, Maryland 21250, USA}
\affiliation{Astroparticle Physics Laboratory, NASA/GSFC, Greenbelt, Maryland 20771, USA}
\affiliation{Center for Research and Exploration in Space Sciences and Technology, NASA/GSFC, Greenbelt, Maryland 20771, USA}
\author{G.~Castellini}
\affiliation{Institute of Applied Physics (IFAC),  National Research Council (CNR), Via Madonna del Piano, 10, 50019, Sesto Fiorentino, Italy}
\author{C.~Checchia}
\affiliation{Department of Physical Sciences, Earth and Environment, University of Siena, via Roma 56, 53100 Siena, Italy}
\affiliation{INFN Sezione di Pisa, Polo Fibonacci, Largo B. Pontecorvo, 3 - 56127 Pisa, Italy}
\author{M.L.~Cherry}
\affiliation{Department of Physics and Astronomy, Louisiana State University, 202 Nicholson Hall, Baton Rouge, Louisiana 70803, USA}
\author{G.~Collazuol}
\affiliation{Department of Physics and Astronomy, University of Padova, Via Marzolo, 8, 35131 Padova, Italy}
\affiliation{INFN Sezione di Padova, Via Marzolo, 8, 35131 Padova, Italy} 
\author{G.A.~de~Nolfo}
\affiliation{Heliospheric Physics Laboratory, NASA/GSFC, Greenbelt, Maryland 20771, USA}
\author{K.~Ebisawa}
\affiliation{Institute of Space and Astronautical Science, Japan Aerospace Exploration Agency, 3-1-1 Yoshinodai, Chuo, Sagamihara, Kanagawa 252-5210, Japan}
\author{A.~W.~Ficklin}
\affiliation{Department of Physics and Astronomy, Louisiana State University, 202 Nicholson Hall, Baton Rouge, Louisiana 70803, USA}
\author{H.~Fuke}
\affiliation{Institute of Space and Astronautical Science, Japan Aerospace Exploration Agency, 3-1-1 Yoshinodai, Chuo, Sagamihara, Kanagawa 252-5210, Japan}
\author{S.~Gonzi}
\affiliation{Department of Physics, University of Florence, Via Sansone, 1 - 50019, Sesto Fiorentino, Italy}
\affiliation{INFN Sezione di Florence, Via Sansone, 1 - 50019, Sesto Fiorentino, Italy}
\affiliation{Institute of Applied Physics (IFAC),  National Research Council (CNR), Via Madonna del Piano, 10, 50019, Sesto Fiorentino, Italy}
\author{T.G.~Guzik}
\affiliation{Department of Physics and Astronomy, Louisiana State University, 202 Nicholson Hall, Baton Rouge, Louisiana 70803, USA}
\author{T.~Hams}
\affiliation{Center for Space Sciences and Technology, University of Maryland, Baltimore County, 1000 Hilltop Circle, Baltimore, Maryland 21250, USA}
\author{K.~Hibino}
\affiliation{Kanagawa University, 3-27-1 Rokkakubashi, Kanagawa, Yokohama, Kanagawa 221-8686, Japan}
\author{M.~Ichimura}
\affiliation{Faculty of Science and Technology, Graduate School of Science and Technology, Hirosaki University, 3, Bunkyo, Hirosaki, Aomori 036-8561, Japan}
\author{K.~Ioka}
\affiliation{Yukawa Institute for Theoretical Physics, Kyoto University, Kitashirakawa Oiwake-cho, Sakyo-ku, Kyoto, 606-8502, Japan}
\author{W.~Ishizaki}
\affiliation{Institute for Cosmic Ray Research, The University of Tokyo, 5-1-5 Kashiwa-no-Ha, Kashiwa, Chiba 277-8582, Japan}
\author{M.H.~Israel}
\affiliation{Department of Physics and McDonnell Center for the Space Sciences, Washington University, One Brookings Drive, St. Louis, Missouri 63130-4899, USA}
\author{K.~Kasahara}
\affiliation{Department of Electronic Information Systems, Shibaura Institute of Technology, 307 Fukasaku, Minuma, Saitama 337-8570, Japan}
\author{J.~Kataoka}
\affiliation{School of Advanced Science and	Engineering, Waseda University, 3-4-1 Okubo, Shinjuku, Tokyo 169-8555, Japan}
\author{R.~Kataoka}
\affiliation{National Institute of Polar Research, 10-3, Midori-cho, Tachikawa, Tokyo 190-8518, Japan}
\author{Y.~Katayose}
\affiliation{Faculty of Engineering, Division of Intelligent Systems Engineering, Yokohama National University, 79-5 Tokiwadai, Hodogaya, Yokohama 240-8501, Japan}
\author{C.~Kato}
\affiliation{Faculty of Science, Shinshu University, 3-1-1 Asahi, Matsumoto, Nagano 390-8621, Japan}
\author{N.~Kawanaka}
\affiliation{Yukawa Institute for Theoretical Physics, Kyoto University, Kitashirakawa Oiwake-cho, Sakyo-ku, Kyoto, 606-8502, Japan}
\author{Y.~Kawakubo}
\affiliation{Department of Physics and Astronomy, Louisiana State University, 202 Nicholson Hall, Baton Rouge, Louisiana 70803, USA}
\author{K.~Kobayashi}
\affiliation{Waseda Research Institute for Science and Engineering, Waseda University, 17 Kikuicho,  Shinjuku, Tokyo 162-0044, Japan}
\affiliation{JEM Utilization Center, Human Spaceflight Technology Directorate, Japan Aerospace Exploration Agency, 2-1-1 Sengen, Tsukuba, Ibaraki 305-8505, Japan}
\author{K.~Kohri} 
\affiliation{Institute of Particle and Nuclear Studies, High Energy Accelerator Research Organization, 1-1 Oho, Tsukuba, Ibaraki, 305-0801, Japan} 
\author{H.S.~Krawczynski}
\affiliation{Department of Physics and McDonnell Center for the Space Sciences, Washington University, One Brookings Drive, St. Louis, Missouri 63130-4899, USA}
\author{J.F.~Krizmanic}
\affiliation{Astroparticle Physics Laboratory, NASA/GSFC, Greenbelt, Maryland 20771, USA}
\author{P.~Maestro}
\affiliation{Department of Physical Sciences, Earth and Environment, University of Siena, via Roma 56, 53100 Siena, Italy}
\affiliation{INFN Sezione di Pisa, Polo Fibonacci, Largo B. Pontecorvo, 3 - 56127 Pisa, Italy}
\author{P.S.~Marrocchesi}
\affiliation{Department of Physical Sciences, Earth and Environment, University of Siena, via Roma 56, 53100 Siena, Italy}
\affiliation{INFN Sezione di Pisa, Polo Fibonacci, Largo B. Pontecorvo, 3 - 56127 Pisa, Italy}
\author{A.M.~Messineo}
\affiliation{University of Pisa, Polo Fibonacci, Largo B. Pontecorvo, 3 - 56127 Pisa, Italy}
\affiliation{INFN Sezione di Pisa, Polo Fibonacci, Largo B. Pontecorvo, 3 - 56127 Pisa, Italy}
\author{J.W.~Mitchell}
\affiliation{Astroparticle Physics Laboratory, NASA/GSFC, Greenbelt, Maryland 20771, USA}
\author{S.~Miyake}
\affiliation{Department of Electrical and Electronic Systems Engineering, National Institute of Technology (KOSEN), Ibaraki College, 866 Nakane, Hitachinaka, Ibaraki 312-8508, Japan}
\author{A.A.~Moiseev}
\affiliation{Department of Astronomy, University of Maryland, College Park, Maryland 20742, USA}
\affiliation{Astroparticle Physics Laboratory, NASA/GSFC, Greenbelt, Maryland 20771, USA}
\affiliation{Center for Research and Exploration in Space Sciences and Technology, NASA/GSFC, Greenbelt, Maryland 20771, USA}
\author{M.~Mori}
\affiliation{Department of Physical Sciences, College of Science and Engineering, Ritsumeikan University, Shiga 525-8577, Japan}
\author{N.~Mori}
\affiliation{INFN Sezione di Florence, Via Sansone, 1 - 50019, Sesto Fiorentino, Italy}
\author{H.M.~Motz}
\email[]{motz@aoni.waseda.jp}
\affiliation{Kanagawa University, 3-27-1 Rokkakubashi, Kanagawa, Yokohama, Kanagawa 221-8686, Japan}
\author{K.~Munakata}
\affiliation{Faculty of Science, Shinshu University, 3-1-1 Asahi, Matsumoto, Nagano 390-8621, Japan}
\author{S.~Nakahira}
\affiliation{Institute of Space and Astronautical Science, Japan Aerospace Exploration Agency, 3-1-1 Yoshinodai, Chuo, Sagamihara, Kanagawa 252-5210, Japan}
\author{J.~Nishimura}
\affiliation{Institute of Space and Astronautical Science, Japan Aerospace Exploration Agency, 3-1-1 Yoshinodai, Chuo, Sagamihara, Kanagawa 252-5210, Japan}
\author{S.~Okuno}
\affiliation{Kanagawa University, 3-27-1 Rokkakubashi, Kanagawa, Yokohama, Kanagawa 221-8686, Japan}
\author{J.F.~Ormes}
\affiliation{Department of Physics and Astronomy, University of Denver, Physics Building, Room 211, 2112 East Wesley Avenue, Denver, Colorado 80208-6900, USA}
\author{S.~Ozawa}
\affiliation{Quantum ICT Advanced Development Center, National Institute of Information and Communications Technology, 4-2-1 Nukui-Kitamachi, Koganei, Tokyo 184-8795, Japan}
\author{L.~Pacini}
\affiliation{INFN Sezione di Florence, Via Sansone, 1 - 50019, Sesto Fiorentino, Italy}
\affiliation{Institute of Applied Physics (IFAC),  National Research Council (CNR), Via Madonna del Piano, 10, 50019, Sesto Fiorentino, Italy}
\author{P.~Papini}
\affiliation{INFN Sezione di Florence, Via Sansone, 1 - 50019, Sesto Fiorentino, Italy}
\author{B.F.~Rauch}
\affiliation{Department of Physics and McDonnell Center for the Space Sciences, Washington University, One Brookings Drive, St. Louis, Missouri 63130-4899, USA}
\author{S.B.~Ricciarini}
\affiliation{INFN Sezione di Florence, Via Sansone, 1 - 50019, Sesto Fiorentino, Italy}
\affiliation{Institute of Applied Physics (IFAC),  National Research Council (CNR), Via Madonna del Piano, 10, 50019, Sesto Fiorentino, Italy}
\author{K.~Sakai}
\affiliation{Center for Space Sciences and Technology, University of Maryland, Baltimore County, 1000 Hilltop Circle, Baltimore, Maryland 21250, USA}
\affiliation{Astroparticle Physics Laboratory, NASA/GSFC, Greenbelt, Maryland 20771, USA}
\affiliation{Center for Research and Exploration in Space Sciences and Technology, NASA/GSFC, Greenbelt, Maryland 20771, USA}
\author{T.~Sakamoto}
\affiliation{College of Science and Engineering, Department of Physics and Mathematics, Aoyama Gakuin University,  5-10-1 Fuchinobe, Chuo, Sagamihara, Kanagawa 252-5258, Japan}
\author{M.~Sasaki}
\affiliation{Department of Astronomy, University of Maryland, College Park, Maryland 20742, USA}
\affiliation{Astroparticle Physics Laboratory, NASA/GSFC, Greenbelt, Maryland 20771, USA}
\affiliation{Center for Research and Exploration in Space Sciences and Technology, NASA/GSFC, Greenbelt, Maryland 20771, USA}
\author{Y.~Shimizu}
\affiliation{Kanagawa University, 3-27-1 Rokkakubashi, Kanagawa, Yokohama, Kanagawa 221-8686, Japan}
\author{A.~Shiomi}
\affiliation{College of Industrial Technology, Nihon University, 1-2-1 Izumi, Narashino, Chiba 275-8575, Japan}
\author{P.~Spillantini}
\affiliation{Department of Physics, University of Florence, Via Sansone, 1 - 50019, Sesto Fiorentino, Italy}
\author{F.~Stolzi}
\affiliation{Department of Physical Sciences, Earth and Environment, University of Siena, via Roma 56, 53100 Siena, Italy}
\affiliation{INFN Sezione di Pisa, Polo Fibonacci, Largo B. Pontecorvo, 3 - 56127 Pisa, Italy}
\author{S.~Sugita}
\affiliation{College of Science and Engineering, Department of Physics and Mathematics, Aoyama Gakuin University,  5-10-1 Fuchinobe, Chuo, Sagamihara, Kanagawa 252-5258, Japan}
\author{A.~Sulaj} 
\affiliation{Department of Physical Sciences, Earth and Environment, University of Siena, via Roma 56, 53100 Siena, Italy}
\affiliation{INFN Sezione di Pisa, Polo Fibonacci, Largo B. Pontecorvo, 3 - 56127 Pisa, Italy}
\author{M.~Takita}
\affiliation{Institute for Cosmic Ray Research, The University of Tokyo, 5-1-5 Kashiwa-no-Ha, Kashiwa, Chiba 277-8582, Japan}
\author{T.~Tamura}
\affiliation{Kanagawa University, 3-27-1 Rokkakubashi, Kanagawa, Yokohama, Kanagawa 221-8686, Japan}
\author{T.~Terasawa}
\affiliation{Institute for Cosmic Ray Research, The University of Tokyo, 5-1-5 Kashiwa-no-Ha, Kashiwa, Chiba 277-8582, Japan}
\author{S.~Torii}
\email[]{torii.shoji@waseda.jp}
\affiliation{Waseda Research Institute for Science and Engineering, Waseda University, 17 Kikuicho,  Shinjuku, Tokyo 162-0044, Japan}
\author{Y.~Tsunesada}
\affiliation{Graduate School of Science, Osaka Metropolitan University, Sugimoto, Sumiyoshi, Osaka 558-8585, Japan }
\affiliation{ Nambu Yoichiro Institute for Theoretical and Experimental Physics, Osaka Metropolitan University,  Sugimoto, Sumiyoshi, Osaka  558-8585, Japan}
\author{Y.~Uchihori}
\affiliation{National Institutes for Quantum and Radiation Science and Technology, 4-9-1 Anagawa, Inage, Chiba 263-8555, Japan}
\author{E.~Vannuccini}
\affiliation{INFN Sezione di Florence, Via Sansone, 1 - 50019, Sesto Fiorentino, Italy}
\author{J.P.~Wefel}
\affiliation{Department of Physics and Astronomy, Louisiana State University, 202 Nicholson Hall, Baton Rouge, Louisiana 70803, USA}
\author{K.~Yamaoka}
\affiliation{Nagoya University, Furo, Chikusa, Nagoya 464-8601, Japan}
\author{S.~Yanagita}
\affiliation{College of Science, Ibaraki University, 2-1-1 Bunkyo, Mito, Ibaraki 310-8512, Japan}
\author{A.~Yoshida}
\affiliation{College of Science and Engineering, Department of Physics and Mathematics, Aoyama Gakuin University,  5-10-1 Fuchinobe, Chuo, Sagamihara, Kanagawa 252-5258, Japan}
\author{K.~Yoshida}
\affiliation{Department of Electronic Information Systems, Shibaura Institute of Technology, 307 Fukasaku, Minuma, Saitama 337-8570, Japan}
\author{W.~V.~Zober}
\affiliation{Department of Physics and McDonnell Center for the Space Sciences, Washington University, One Brookings Drive, St. Louis, Missouri 63130-4899, USA}

\collaboration{CALET Collaboration}

\date{\today}

\begin{abstract}
 Detailed measurements of the spectral structure of cosmic-ray electrons and positrons from 10.6~GeV to 7.5~TeV are presented
from over 7 years of observations with the CALorimetric Electron Telescope (CALET) on the International Space Station.   The instrument, consisting of a charge detector, an imaging calorimeter, and a total absorption calorimeter with a total  depth of 30 radiation lengths at normal incidence and a fine shower imaging capability, is optimized to measure the all-electron spectrum well into the TeV region. Because of the excellent energy resolution (a few percent above 10~GeV) and the outstanding e/p separation ($10^5$), CALET provides optimal performance for a detailed search of structures in the energy spectrum.
 The analysis uses data up to the end of 2022, and the statistics of observed electron candidates has increased more than 3 times since the last publication in 2018.  By adopting an updated boosted decision tree analysis, a sufficient proton rejection power up to 7.5~TeV is achieved, with a residual proton contamination less than 10\%.
 The observed energy spectrum  becomes gradually harder in the lower energy region from around 30~GeV, consistently with AMS-02, but from 300 to 600~GeV it is considerably softer than the spectra measured by DAMPE and Fermi-LAT.
  At high energies, the spectrum presents a sharp break around 1~TeV, with a spectral index change from -3.15 to -3.91, and a broken power law fitting the data in the energy range from 30~GeV to 4.8~TeV better than a single power law with 6.9 sigma significance, which is compatible with the DAMPE results.
The break is consistent with the expected effects of radiation loss during the propagation from  distant sources (except the highest energy bin).  
 We  have fitted the spectrum with a model consistent with the positron flux measured by AMS-02 below 1~TeV and interpreted the electron + positron spectrum with possible contributions from  pulsars and nearby sources. Above 4.8~TeV, a possible contribution from known nearby supernova remnants, including Vela, is addressed  by  an event-by-event analysis providing a higher proton-rejection power than a purely statistical analysis. 

 \end{abstract}


\maketitle

{\it Introduction.\,---}
Direct measurements of high-energy electron and positron cosmic rays (hereafter, all-electron CRs)  have advanced significantly since  the 2000s with state-of-the art detectors in space, some of which continue to operate increasing the collected statistics and, correspondingly, the precision of the spectrum. Based on  these observations, it has  widely been recognized  that the all-electron  spectrum cannot be described by a single power law  in the range from the 10~GeV to the TeV region. In particular,  the energy spectrum above 1~TeV is expected to show a break due to the radiative cooling process with an energy loss rate proportional to $E^2$. As a result, only nearby ( $<$ 1~kpc) and young ( $< 10^5$ yr) sources can contribute to the flux above 1~TeV if the sources are supernova remnants (SNRs) as it is commonly believed. 
  The pioneering works~\cite{SH1970, CL1979, JN1980, AAV1995, PE1998, EW2002, TK2004} pointed out a possible break of the electron spectrum above 1 TeV, suggesting that precise measurements of the spectrum in the TeV region could lead to the identification of nearby sources. 
  Recently, several authors interpreted the observed spectral break above 1~TeV assuming this scenario (for example, Refs.~\cite{PL2019,OF2020,YD2021,KA2022}. Also, a direct probe of the escape mechanism from SNR is discussed, for example, in Ref.~\cite{NK2011}.
\par
The calorimetric electron telescope (CALET) is a space experiment installed at the Japanese Experiment Module--Exposed Facility (JEM--EF) on the International Space Station (ISS) for long term observations of cosmic rays and optimized for the measurement of the all-electron spectrum~\cite{CALET-overview} . 
The first result on the all-electron spectrum by CALET was published in the energy range from 10~GeV to 3~TeV, the first ever significant observation reaching into the TeV region~\cite{CALET-ae1}. Subsequently, an updated spectrum was published with a factor $\sim$~2 larger statistics   by using more than 2 years of flight data and the full geometrical acceptance in the high-energy region~\cite{CALET-ae2}.
The observed energy spectrum above $\sim$1~TeV suggests a flux suppression consistent within the errors with the results of dark matter particle explore (DAMPE)~\cite{DAMPE-ae}.
\\
\hspace{2ex}
Although calorimeters as CALET and the DAMPE~\cite{DAMPE-chang} are not able to measure the polarity of charge, magnet spectrometers, such as the  payload for antimatter matter exploration and light nuclei astrophysics (PAMELA)~\cite{PAMELA-pe} and the alpha magnetic spectrometer (AMS-02)~\cite{AMS-pe}, measured separately the positrons and the electrons, and found an increase of the positron fraction above 10~GeV.   The fraction reaches a maximum ($\sim$ 15 \%) around 300~GeV and decreases to a level of a few percent near 1~TeV. The results  require a primary source component of the positrons in addition to the generally accepted secondary origin. Candidates for such primary sources range from astrophysical (pulsar) to exotic (dark matter). Since these primary sources emit electron-positron pairs, it is expected that the shape of the all-electron spectrum would reflect the presence of  the primary source component of electrons and positrons, in the corresponding energy range above 10~GeV.
 
In this paper, we present the CALET all-electron spectrum with a further increase in statistics by a factor $\sim$~3.4 since the last publication~\cite{CALET-ae2}, using 2637 days of flight data from October 13, 2015 to December 31, 2022.
The spectrum integrates 7.02 million electron (+ positron) events above 10.6~GeV up to 7.5~TeV.
Combining the CALET all-electron spectrum and the positron measurements up to 1 TeV by AMS-02, we attempt a  consistent interpretation of   both spectra based on contributions from  pulsars and nearby SNR sources.
Based on this interpretation, the obtained spectrum in the TeV region is tested for indications of contributions from the nearby sources, foremost Vela, by estimating the number of electron candidates above 4.8~TeV obtained with an event-by-event analysis with a residual proton contamination probability less than 10\%~\cite{CALET-ae3SM}.

{\it Instrument.\,---}
CALET is a fully active calorimeter optimized for electron observations  from 1~GeV up to 20~TeV.
It consists of a charge detector (CHD), a 3 radiation-length thick imaging calorimeter (IMC), and a 27 radiation-length thick total absorption calorimeter (TASC). It has a field of view of approximately 45$^\circ$ from zenith and a geometrical factor of 1040~cm$^2$ sr for high-energy electrons.
The IMC induces the start of the shower development for electromagnetic particles while suppressing nuclear interactions in order to maximize the proton rejection power for the electron candidates, and  provides the direction of incident particles.
It  is composed of  7  layers of  tungsten absorbers interleaved with scintillating fiber belts read out individually with 64-anode PMTs.  The TASC  installed below the IMC measures the energy of shower particles caused by the interactions of the incident particles in the IMC.  It is a tightly packed lead-tungstate (PbWO$_4$; PWO) hodoscope, allowing for a nearly total  containment of TeV-electron showers. The CHD, placed at the top of the detector to identify the charge of the incident particle, is comprised of a pair of plastic scintillator hodoscopes arranged in two orthogonal layers.  

With the precise energy measurements from total absorption of electromagnetic showers, it is possible to derive the electron spectrum well into the TeV region with a straightforward and reliable analysis. A more complete description of the instrument is given in Ref.~\cite{CALET-ae1SM}.

%

{\it Observation and calibrations.\,---}
 Since the start of  scientific operations,  CALET observations have been carried out continuously without any serious incident and  with downtime less than a few days during each interruption. The live time fraction, dominated by the data  acquisition dead time (nearly 5 ms per event)  is nearly  86 \%,  including runs for calibration and the high trigger rate for low energy particles ($>$1~GeV) ~\cite{CALET-obs}. 
 The total live time was  so $1.927 \times 10^8$ sec. 

CALET carries out precise  energy measurements over a very wide dynamic range from 1~GeV to 1~PeV by exploiting the read-out system of the TASC, which implements  four gain  ranges for each channel, providing excellent energy resolution even in the TeV region.  Our energy calibration includes the evaluation of the conversion factors between analog-to-digital converter units and energy deposits, ensuring linearity over each gain range  and provides a seamless transition between neighboring gain ranges~\cite{CALET-calib}. The absolute calibration of energy is done by using the energy deposit of penetrating protons and/or helium particles detected at the highest gain. 

Temporal gain variations occurring during long duration observations are also corrected by the calibration procedure. The errors at each calibration step, such as the correction of position and temperature dependence, consistency between energy deposit peaks of noninteracting protons and helium, linear fit error of each gain range, and gain ratio measurements, as well as slope extrapolation, are included in the estimation of the energy resolution.  As a result, an excellent energy resolution of 2\% or better is achieved above 20~GeV up to over 10~TeV.
The calibrations are checked monthly to confirm the instrument stability, and the spectra of  deposited energies in TASC  using four gain ranges are compared among each other for consistency. 
 
{\it Data analysis.\,---}
The analysis has been carried out  on the data collected with a high-energy shower trigger~\cite{CALET-obs} in the full detector acceptance, by an updated procedure to reduce  the proton background in the TeV region, compared with  the analysis described in Ref.~\cite{CALET-ae2}. A Monte Carlo (MC) program was used to simulate physics processes and detector response based on the simulation package EPICS~\cite{EPICS} (EPICS9.20/COSMOS8.00). Using MC event samples of electrons and protons, event selection and event reconstruction efficiencies, energy correction factors, and background contamination were derived. An independent analysis based on GEANT4~\cite{GEANT} was performed, and  differences between the MC models are included in the systematic uncertainties. The GEANT4 simulation employs the hadronic interaction models FTFP-BERT as the physics list, while DPMJET3~\cite{DPMJET3} is chosen as the hadronic interaction model in the EPICS simulation.

We use the "electromagnetic shower tracking" algorithm~\cite{emtrack} to reconstruct the shower axis of each event, taking advantage of the electromagnetic shower shape and IMC imaging capabilities. As input for the electron identification, well-reconstructed and well-contained single-charged events are preselected by (i) an offline trigger confirmation, (ii) a geometrical condition, (iii) a track quality cut to ensure reconstruction accuracy, (iv) a charge selection using CHD, (v)  a requirement based on the longitudinal shower development, and (vi) on the lateral shower consistency with that expected for electromagnetic cascades.

In addition to fully contained events, the events incident from the IMC sides and exiting through the sides of TASC are used for analysis above 
476~GeV~\cite{CALET-ae2}.  For events not crossing the CHD, we use the energy deposit of the first hit IMC layer to determine their charge. The path length inside TASC is required to be longer than the vertical depth of TASC, i.e., 27 radiation lengths. The energy of incident electrons is reconstructed using an energy correction function which converts the energy deposit  of TASC and IMC into primary energy for each geometrical condition.The absolute energy scale was calibrated and shifted by +3.5\% ~\cite{CALET-ae1} as a result of a study of the geomagnetic cutoff. Since the full dynamic range calibration~\cite{CALET-calib} was carried out with a scale-free method, its validity holds regardless of the absolute scale uncertainty. The systematic uncertainties are described in detail in  the Supplemental Material~\cite{CALET-ae3SM} .

 In order to identify electrons and to study systematic uncertainties in the electron identification, we applied two methods: a simple two-parameter cut below 476~GeV and a multivariate analysis above. The latter is based on boosted decision trees (BDTs) optimized in the  energy interval above (below)  949~GeV, using 13 ~(9) parameters, respectively.   Calculation of event selection efficiencies, BDT training, and estimation of proton background contamination are carried out separately for each geometrical condition and combined in the end to obtain the final spectrum. Considering  that the lower energy region is dominated by systematics in our analysis, and therefore more statistics would not significantly improve the precision of our data, only fully contained events are included in the lower energy region below 476~GeV. 

An example of a BDT response distribution in the 754 $< E <$ 949~GeV bin including all acceptance conditions is shown in Fig.~\ref{fig:BDT}. The BDT response distributions for the TeV region are shown in Fig.~S1 of the Supplemental Material~\cite{CALET-ae3SM}. In the final electron sample, the  contamination ratios of protons are 5\% up to 1~TeV,  and less than 10\% in the 1--7.5~TeV region, while keeping a constant high efficiency of 80\% for electrons.
\begin{figure}[th!]
\begin{center}
\includegraphics[width=1.0 \linewidth]{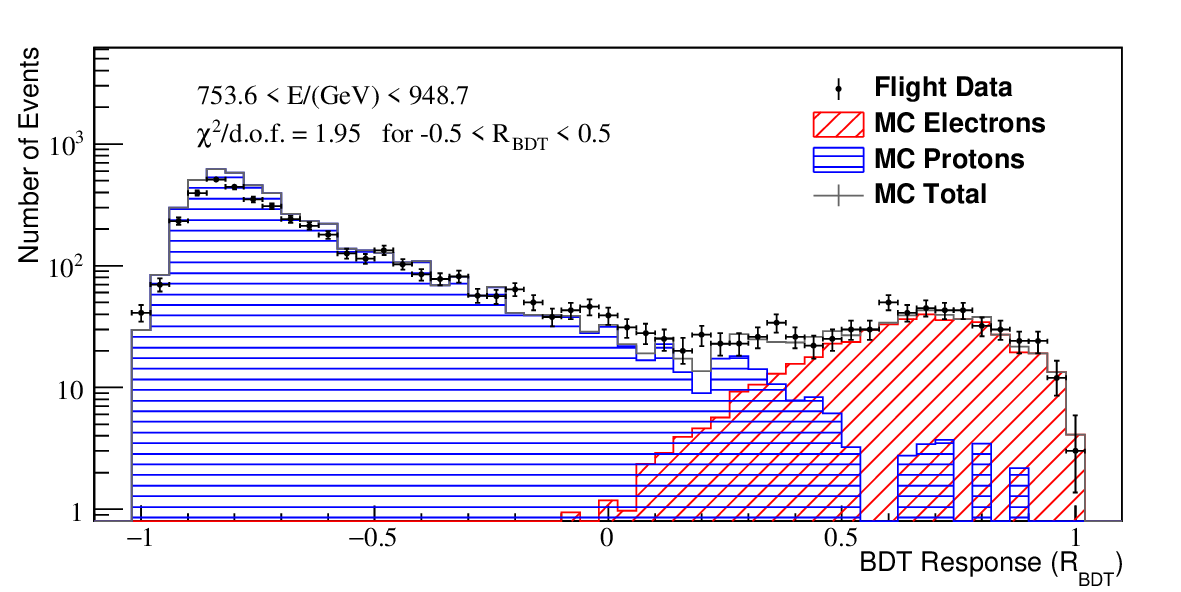}
\caption{An example of BDT response distributions in the 754 $< E <$ 949~GeV bin, including all acceptance conditions. }
\label{fig:BDT}
\end{center}
\end{figure}

By using the data obtained with the low energy trigger (1~GeV threshold), the high energy trigger efficiency was verified, considering only the events observed   in the rigidity cutoff region below 6~GV.
Two independent  analyses were carried out by separate groups inside the CALET Collaboration, using different event selections and acceptance of the event geometries. The results of the two analyses are consistent with each other within the errors  over the entire energy region.

\begin{figure}[h!]
\begin{center}
\includegraphics[width=1.0 \linewidth]{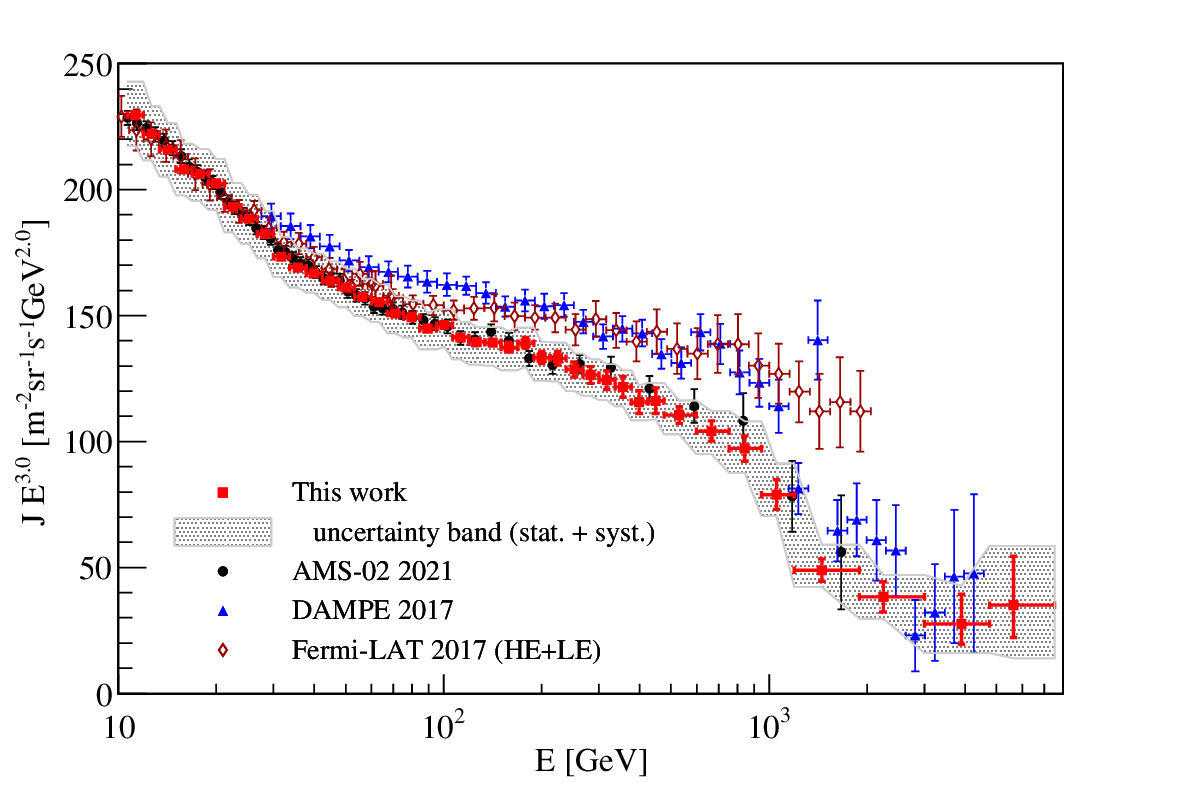}
\caption{Cosmic-ray all-electron spectrum measured by CALET from 10.6~GeV to 7.5~TeV using the same
energy binning as in our previous publication below 4.8~TeV~\cite{CALET-ae2}, where the gray band indicates the quadratic sum of
statistical and systematic errors (not including the uncertainty on the energy scale). Also plotted are other direct
measurements in space~\cite{AMS-ae21, Fermi-ae, DAMPE-ae} for comparison. The enlarged figure is shown in Fig.~S4 in the Supplemental Material~\cite{CALET-ae3SM}.}
\label{fig:Spectrum}
\end{center}
\end{figure}

{\it Results.\,---}
Figure~\ref{fig:Spectrum} shows the all electron spectrum obtained in this analysis using the observed events with statistics increased by a factor 3.4 since the last publication ~\cite{CALET-ae2}. The error bars along the horizontal and vertical axes indicate the bin width and statistical errors, respectively. The gray band is representative of the quadratic sum of statistical and systematic errors, using the same definition as in Ref. ~\cite{CALET-ae2}. 

Systematic errors include errors in the absolute normalization and energy dependent ones.
The energy dependent errors include those obtained from BDT stability, trigger efficiency in the low-energy region, tracking dependence, dependence on methods of charge identification and of electron identification, as well as MC model dependence.
 Conservatively, all of them are included in the total error estimate in Fig.~\ref{fig:Spectrum},  and a breakdown of the contributions from each source and their specific energy dependence is given in the Supplemental Material~\cite{CALET-ae3SM}.  Utilizing this additional data,  our all-electron spectrum in combination with the positron-only measurement by AMS-02 can provide essential information for investigating spectral features as possible signatures of dark matter and/or astrophysical sources. 

Comparing with other recent experiments in space (AMS-02, Fermi-LAT, and DAMPE), the CALET spectrum shows good agreement with AMS-02 data up to 2~TeV. In the energy region from 30 to 300~GeV, the fitted power-law spectral  index, -3.14 $\pm$0.02, is roughly consistent with the values quoted by other experiments  within errors. However, the CALET spectrum appears to be softer   compared to DAMPE and Fermi-LAT, and the flux measured by CALET is lower than that seen by DAMPE and Fermi-LAT, starting near 60~GeV and extending to near 1~TeV, indicating the presence of unknown systematic effects. Moreover, the flux in the 1.4~TeV bin of DAMPE's spectrum,
which might imply a peak structure, is not compatible with CALET results at a significance level of 4.8~$\sigma$
using the same energy binning as DAMPE, including all systematic errors from both experiments.
In Fig.~S5~\cite{CALET-ae3SM}, we show the CALET all-electron spectrum in DAMPE's binning for reference.

\begin{figure}[htb!]
\begin{center}
\includegraphics[width=0.85 \linewidth]{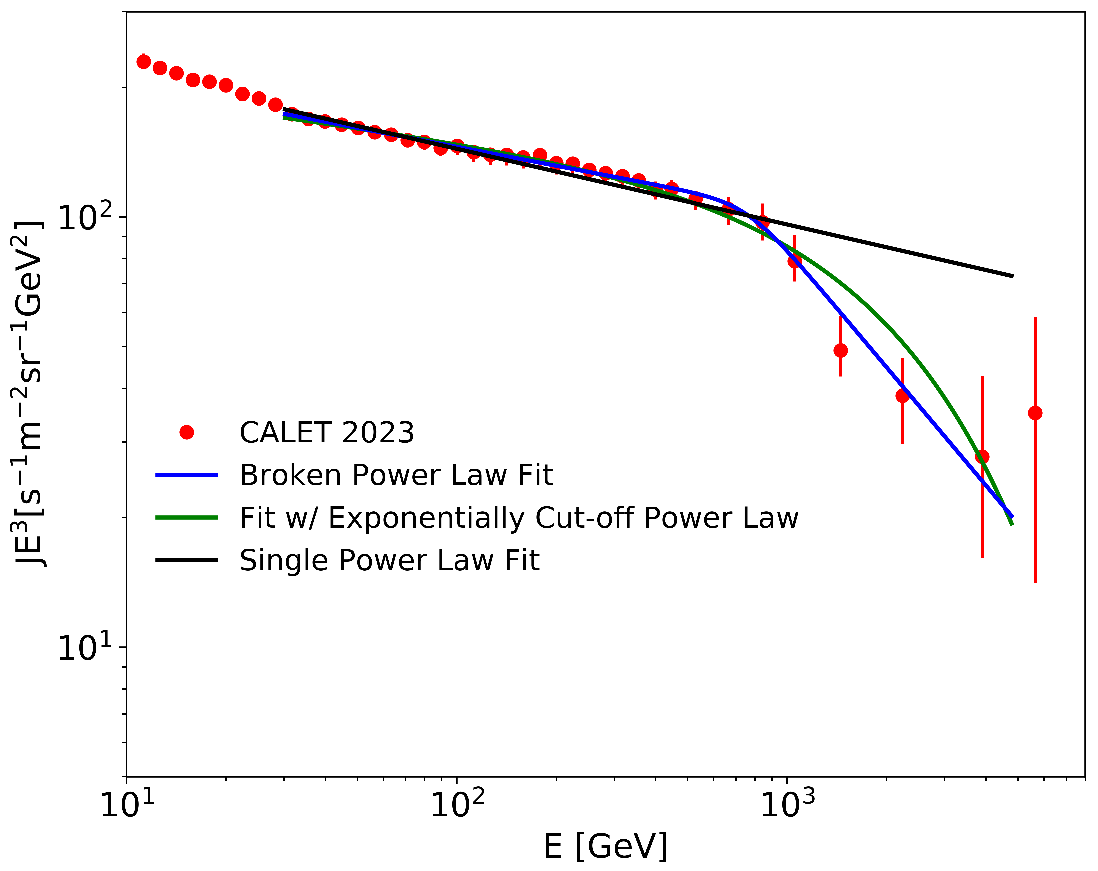}
\caption{All-electron spectrum measured by CALET from 10.6~GeV to 7.5~TeV, and the fitted results in the energy range from 30~GeV to 4.8~TeV, with a broken power law, an exponentially cutoff power law and a single power law. The error bars represent statistical and systematic uncertainties except normalization. See text for the details of the fits by power laws.}
\label{fig:FitSpectrum}
\end{center}
\end{figure}

In Fig.~\ref{fig:FitSpectrum}, we fit the differential spectrum in the energy range from 30~GeV to 4.8~TeV with a smoothly broken power-law model (blue line) ~\cite{SBPLM}. 
The model is defined as: \( J(E)=C(E/100~\mathrm{GeV})^\gamma (1+(E/E_b)^{\Delta\gamma/s} )^{-s} \), where $E_b$ is the break energy, while $\gamma $ is the power index below $E_b$ and $\Delta\gamma$ is the difference in the  power index below and above  $E_b$. The fitted spectrum steepens from 
$\gamma =-3.15 \pm 0.01$  by $\Delta\gamma = -0.77 \pm 0.22$ at energy $E_b = 761 \pm 115$~GeV with the break smoothness parameter (s) fixed to 0.1 which fits our data well, with $\chi^2$ = 3.6 and 27 degrees of freedom (NDF). 

 A single power-law fit over the same energy range (black line) gives $\gamma $=-3.18$\pm$0.01 with $\chi^2$/NDF=56/29, which means that a broken power law is favored with 6.9~$\sigma$ significance over a single power law. An exponentially cutoff power law~\cite{Fermi-ae} (green line) with a  power index of -3.10$\pm$0.01 below a cutoff energy of 2854$\pm$305~GeV fits also our data well, with $\chi^2$/NDF=12/28 and a significance of 6.6$\sigma$ over the single power law.

\begin{figure}[t!]
\begin{center}
\includegraphics[width=0.9\linewidth]{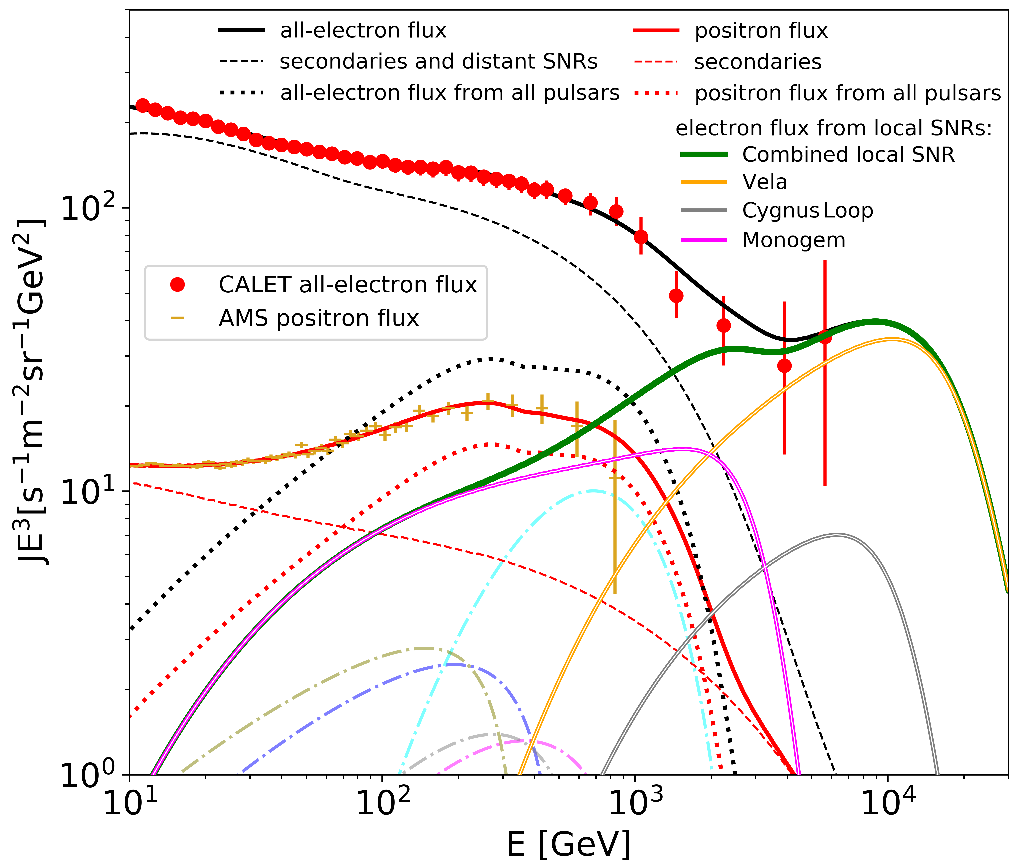}
\caption{Possible spectral fit over the whole region of CALET observations, including pulsars and nearby SNR sources as individual sources, with the Vela SNR dominating in the TeV region. See details in text.}
\label{fig:VelaSpectrum}
\end{center}
\end{figure}

%
{\it Discussion.\,---}
In the following we discuss a possible interpretation of the CALET energy spectrum over the whole energy range. We have incorporated the measured AMS-02 positron flux~\cite{AMS-ae21}, source and propagation parameters suggested in Ref.~\cite{Motz-ICRC2021}, and results from the numerical propagation code DRAGON~\cite{DRAGON} to construct a possible model that fits the CALET all-electron measurements. Figure~\ref{fig:VelaSpectrum} shows the prediction of our example model compared to the CALET results. The positron flux of AMS-02 is fitted with contributions from secondaries (red dashed line) + several pulsars (red dotted line), while the all-electron flux is fitted with the sum of electron and positron flux from the pulsars (black dotted line), in addition to secondaries + distant SNRs (black dashed line) with a cutoff at 1~TeV.
In this model we follow a hypothesis that the positron excess is caused by a primary source of \el+\ps~pairs, for which we include the only contribution from pulsars neglecting more exotic sources as dark matter.
In the range from about 30~GeV to 1~TeV, this \el+\ps~pair source significantly influences the all-electron spectrum. Above 1~TeV, we include the nearby SNRs, Vela (orange solid line), Cygnus Loop (gray solid line) and Monogem (magenta solid line) as the dominant sources~\cite{TK2004}, with their combined contribution (green line). The best fit yields an energy output of 0.8 $\times 10^{48}$ erg in electron cosmic rays above 1~GeV for each nearby SNR.

 The spectra of the nearby SNRs and secondaries (\el, \ps) are calculated using DRAGON~\cite{DRAGON}, which is also used to define the propagation parameters via calculation of the nuclei spectra, concurrently providing spectra of the secondary electrons and positrons forming part of the background. 
This whole-region model for the interpretation of the all-electron spectrum and its implications for the possible contribution of nearby sources is discussed in more detail in Ref.~\cite{Motz-ICRC2021}.
 For the fitting shown in Fig.~\ref{fig:VelaSpectrum}, statistical and systematic errors are added up quadratically, the cutoff energy for the near SNR source spectrum is 100~TeV, and the propagation conditions labeled as  “Model X” in Ref.~\cite{Motz-ICRC2021} are used.  
 The predicted number of events with the best fit is 11.0~(4.2) electrons above 4.8~TeV~(7.5~TeV).
 A fit of the model without the three nearby SNRs and a smooth extension of the power-law spectrum to the TeV-region (Fig.~S6~\cite{CALET-ae3SM} has similar fit quality and predicts 4.6 (1.0) events.
The observed numbers of  electron candidates  obtained by the event-by-event analysis are  9~(4) above 4.8~TeV~(7.5~TeV) , compatible with the expected contribution from the nearby SNRs. A study on the significance while taking  the errors into account will be published  elsewhere.
 The electron selection above 4.8~TeV using an event-by-event analysis is  discussed  in detail in the Supplemental Material~\cite{CALET-ae3SM}.   

{\it Conclusion.\,---}
We have extended our previous result ~\cite{CALET-ae2} of the CALET all-electron spectrum with an approximate increase of the statistics by a factor 3.  The all-electron energy spectrum over the entire region is fitted using the positron flux measured by AMS-02 and the expected contribution of the known astrophysical sources including nearby pulsars and SNRs. 
In the TeV region the data show a break of the spectrum compatible with the DAMPE results. The accuracy of determining the  break's sharpness and
position, and of the spectral shape above 1~TeV, are improved by the better statistics. The observed 9 electron candidates above 4.8~TeV are consistent with an estimation of the electron flux from the nearby SNRs based on an interpretation  model.  Further observations are needed to reach a final conclusion. 

Extended CALET operations approved by JAXA/NASA/ASI in March 2021 through the end of 2024 (at least)
 will bring a further increase of the statistics and a reduction of the systematic errors based on the analysis.\\
\newline
\newline
\indent

We gratefully acknowledge JAXA's contributions to the development of CALET and to the operations onboard the
International Space Station. We also express our sincere gratitude to ASI and NASA for their support of the CALET project. This work was supported in part by JSPS Grant-in-Aid for Scientific Research (S) Grant No.~19H05608, JSPS Grant-in-Aid for Scientific Research (C) Grand No.~21K03592, No.~21K03604, and by the MEXT-Supported Program for the Strategic Research Foundation at Private Universities (2011--2015) (Grant No.~S1101021) at Waseda University. The CALET effort in Italy is supported by ASI under Agreement No.~2013-018-R.0 and its amendments. The CALET effort in the U.S. is supported by NASA through Grants No.~80NSSC20K0397, No.~80NSSC20K0399, and
No.~NNH18ZDA001N-APRA18-004, and under Grant 384 No.~80GSFC21M0002.
 
\end{document}


\widetext
\clearpage
\begin{center}
\end{center}
\setcounter{equation}{0}
\setcounter{figure}{0}
\setcounter{table}{0}
\setcounter{page}{1}
\makeatletter
\renewcommand{\theequation}{S\arabic{equation}}
\renewcommand{\thefigure}{S\arabic{figure}}
\renewcommand{\bibnumfmt}[1]{[S#1]}
\renewcommand{\citenumfont}[1]{S#1}

\begin{center}
\textbf{\Large Direct Measurement of the Spectral Structure of Cosmic-Ray Electrons+Positrons 
  in the TeV Region with CALET  on the International Space Station}
   \end{center}
\vspace*{0.5cm}
Supplemental material relative to "Direct Measurement of the Spectral Structure of Cosmic-Ray Electrons+Positrons 
  in the TeV region with CALET  on the International Space Station''
\clearpage
\begin{center}
{\bf BDT RESPONSE DISTRIBUTION}
\end{center}

\begin{figure}[h!] 
\begin{center}
\includegraphics[width=0.9\linewidth]{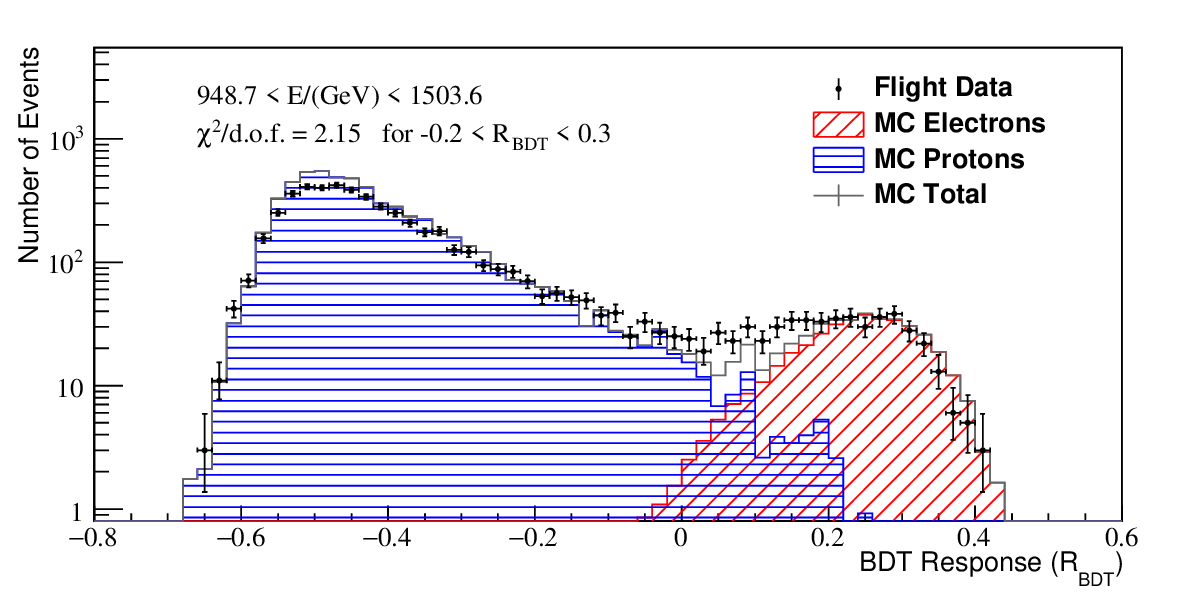}
%
%
\includegraphics[width=0.9 \linewidth]{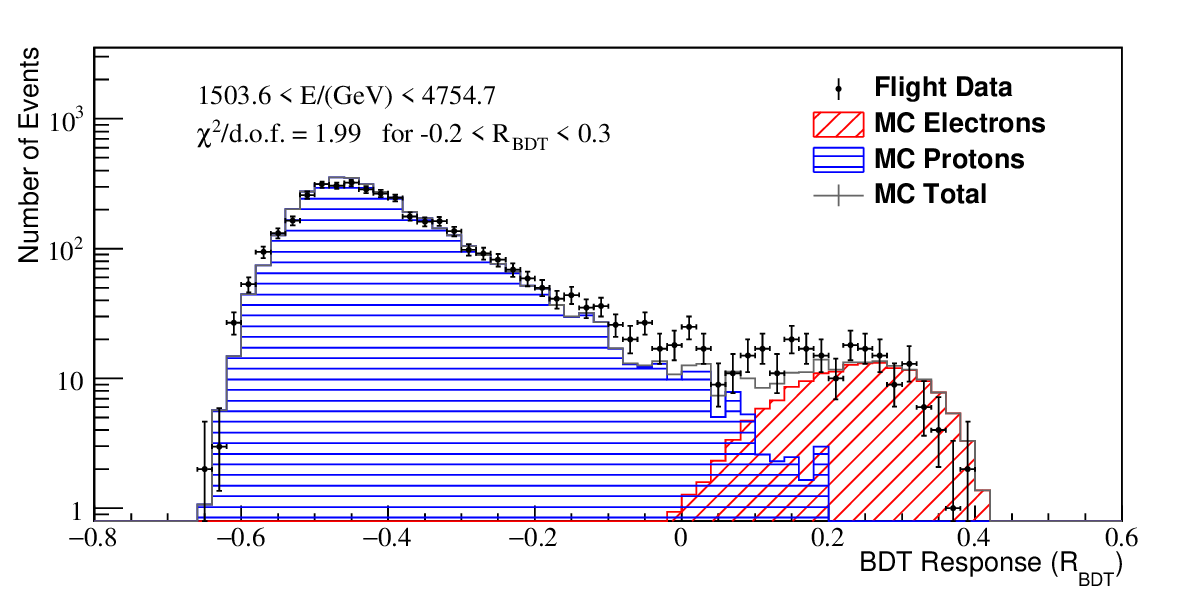}
\caption{BDT distribution in 949 $< E <$ 1504~GeV (top) and 1504 $< E <$ 4755~GeV(bottom) bins.}
\label{fig:BDTdistribution}
\end{center}
\end{figure}

Figure ~\ref{fig:BDTdistribution} presents the BDT response ($R_{BDT}$) distributions in the 949 $< E <$ 1504~GeV (top) and 1504 $< E <$ 4755~GeV
(bottom) bins including all acceptance conditions~\cite{CALET-ae2SM}. While there are noticeable discrepancies in  the -0.2  $<$ $R_{BDT}$ $<$ 0.3 region, their possible effects to the resultant spectrum are included in the systematic uncertainty relative to the BDT stability.
%

\begin{figure}[h!] 
\begin{center}
\includegraphics[width=0.75 \linewidth]{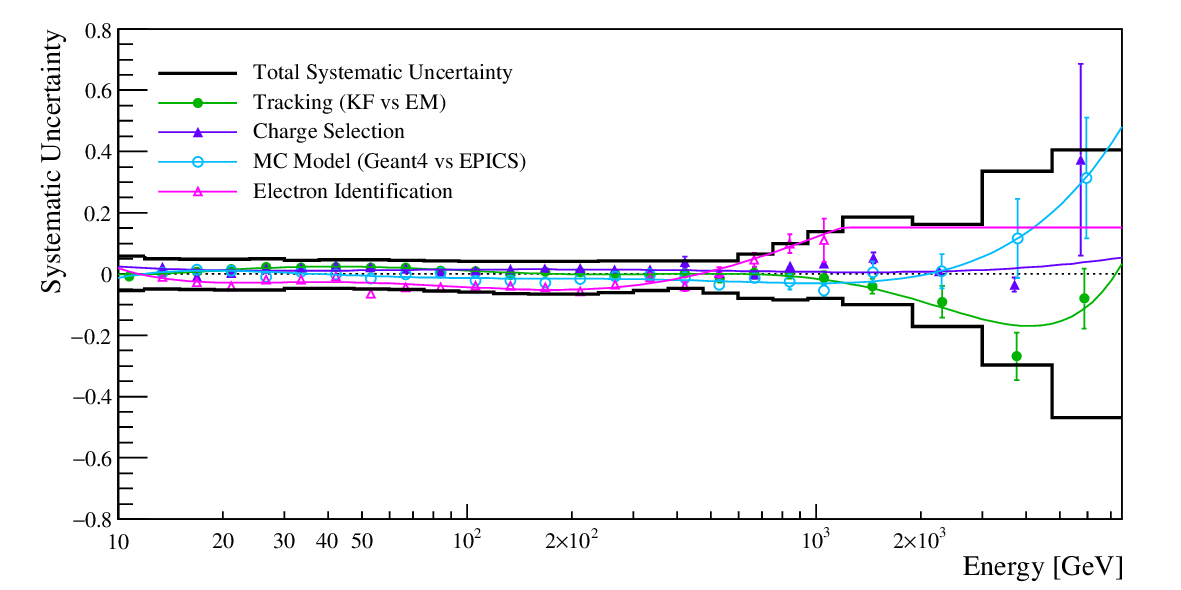}
\caption{Energy dependence of systematic uncertainties. The solid line represents the total systematic. See text for details.}
\label{fig:SystSummary}
\end{center}
\end{figure}

\clearpage
\begin{center}
{\bf SYSTEMATIC UNCERTAINTIES}
\end{center}
Figure~\ref{fig:SystSummary} shows the energy dependence of systematic uncertainties in the tracking algorithms (Electromagnetic shower tracking vs
combinatorial Kalman filter tracking), charge selction methods (CHD vs IMC), 
electron identification methods (K-estimator vs BDT) and MC models (Geant4 vs EPICS).
The data points are fitted with log-polynomial functions to mitigate the effect of statistical fluctuations while preserving possible energy dependent structures. Fit functions are shown as curves and are used to estimate energy dependent systematic uncertainties.
These resultant functions are used to quote the systematic uncertainty considering the sign of the difference between implemented and comparison method in each energy bin.
The K-estimator is not appropriate at high energies due to the increasing background, therefore the function is extrapolated as constant maximum above the fit range.

\begin{figure}[h!] 
\begin{center}
\includegraphics[width=0.75 \linewidth]{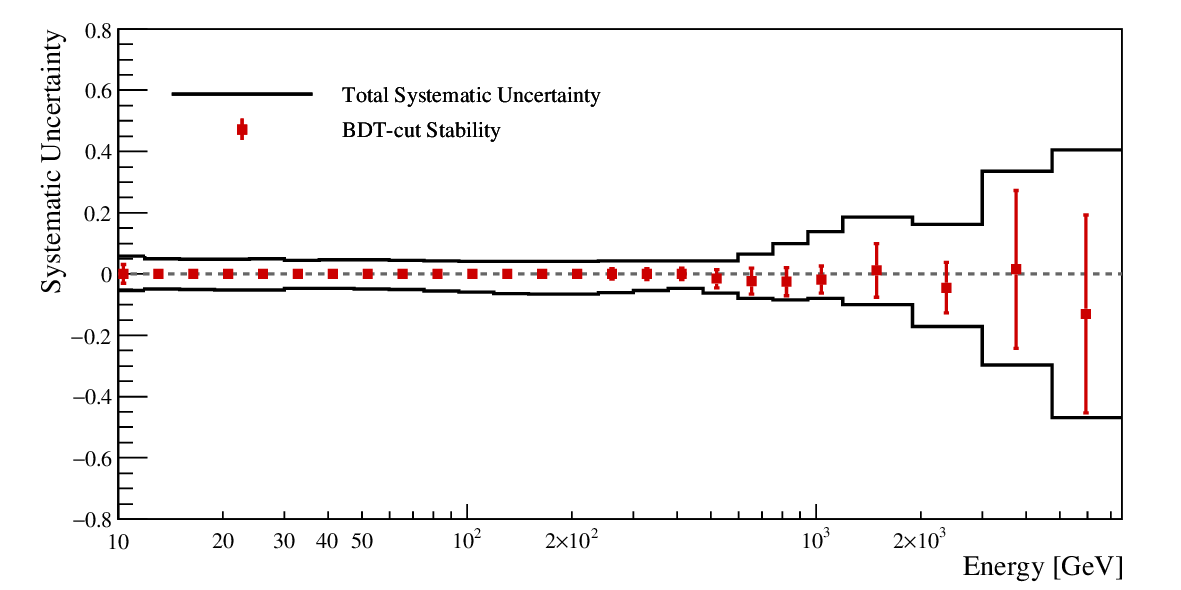}
\caption{Energy dependence of systematic uncertainties. The red squares represent the average of all BDT training samples with respect to the standard 80\% efficiency case, while error bars represent the standard deviation at each energy bin. The upper and lower edge of the error bars are taken as the total systematic uncertainty due to the BDT analysis.}
\label{fig:SystBDT}
\end{center}
\end{figure}

\begin{figure}[h!] 
\begin{center}
\includegraphics[width=1.0 \linewidth]{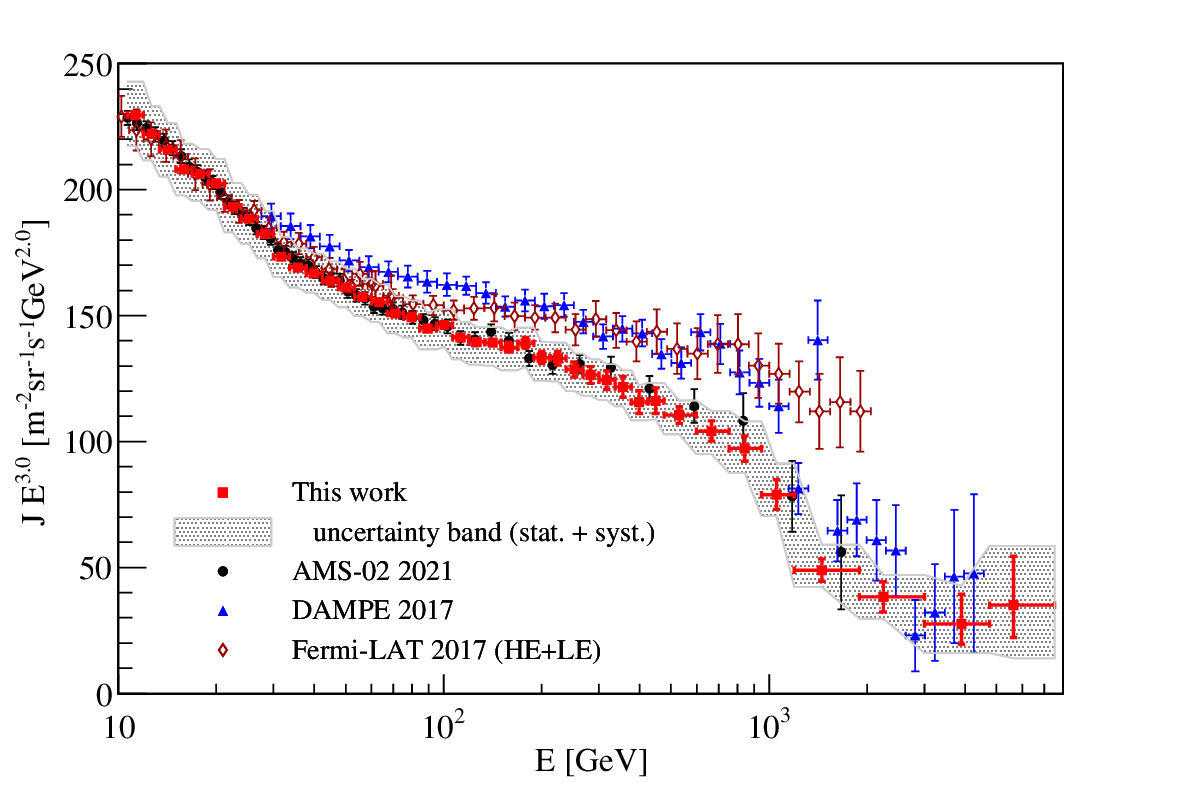}
\caption{Cosmic-ray electron + positron spectrum observed with CALET from 10.6~GeV to~7.5 TeV, compared with other direct measurements. The horizontal error bars are representative of the bin width. }
\label{fig:CALET_Electron}
\end{center}
\end{figure}

BDT stability is evaluated from the stability of the flux obtained with 100 independent training samples as a function of BDT cut efficiency in the range from 70\% to 90\% in 1\% steps for each corresponding test sample.
The energy dependence of the BDT stability is shown in Fig.~\ref{fig:SystBDT}.
\clearpage
\begin{center}
{\bf ENLARGED FIGURE OF ALL-ELCTRON SPECTUM OF CALET}
\end{center}
The electron and positron spectrum measured by CALET (red circles) is shown in Fig.~\ref{fig:CALET_Electron} 
where it is compared with the experimental results of AMS-02~\cite{AMS-ae},  Fermi-LAT~\cite{Fermi-ae}, and DAMPE~\cite{DAMPE-ae}. The hatched band shows the total uncertainty for CALET. 
\clearpage
%
\renewcommand{\arraystretch}{1.3}
\begin{table}
\caption{
Table of CALET electron plus positron spectrum. Mean energy is calculated using the candidate events in the
energy bin. For the flux,  the first and second errors represent the statistical uncertainties (68 \% confidence level) and systematic
uncertainties, respectively.
Detailed breakdown of systematic errors is included where 
$\sigma_{\rm{BDT}}$, $\sigma_{\rm{trig.}}$, $\Delta_{\rm{norm.}}$, $\Delta_{\rm{trk.}}$, $\Delta_{\rm{chg.}}$, $\Delta_{\rm{ID}}$  and 
$\Delta_{\rm{MC}}$ denote systematic errors due to BDT stability, trigger, absolute normalization, tracking, charge identification, electron identification, and MC model dependence, respectively. While the first two components must be added in quadrature to the statistical errors in a spectral analysis, the latter five contributions could be treated by introducing weight factors corresponding to each component as fitted nuisance parameters. This constrains the possible correction to the fit function from each component to a shift with the already determined energy dependence (or non-dependence). The likelihood of the correction from each component enters into the fit quality by adding the squared weight factor of each component to the $\chi^2$. 
Although $\Delta_{\rm{norm.}}$ can be ignored in a spectral study using only CALET data, such as the fits shown in Figs.~3 and S5 , it should also be treated as a nuisance parameter in a combined analysis with the positron spectrum.
}
\label{tab:e3flux}
 \begin{center}
\begin{tabular}[c]{cccccccccc} \hline \hline 
Energy Bin & Mean Energy & Flux & \multicolumn{7}{c}{Systematic Uncertainties (relative to flux)} \\
(GeV) & (GeV) & ($\rm{m^{-2}sr^{-1}s^{-1}GeV^{-1}}$) & $\sigma_{\rm{BDT}}$ & $\sigma_{\rm{trig.}}$ & $\Delta_{\rm{norm.}}$ & $\Delta_{\rm{trk.}}$ & $\Delta_{\rm{chg.}}$ & $\Delta_{\rm{ID}}$ & $\Delta_{\rm{MC}}$ \\
\hline
10.6--11.9 & $11.3$ &$(1.599 \, \pm 0.002 \, _{-0.085}^{+0.091}) \times 10^{-1}$ & 0.031 & 0.024 & 0.036 & -0.001 & 0.021 & 0.002 & -0.003\\
11.9--13.4 & $12.6$ &$(1.099 \, \pm 0.002 \, _{-0.052}^{+0.055}) \times 10^{-1}$ & 0.015 & 0.024 & 0.036 & 0.000 & 0.018 & -0.009 & 0.004\\
13.4--15.0 & $14.2$ &$(7.575 \, \pm 0.013 \, _{-0.373}^{+0.374}) \times 10^{-2}$ & 0.015 & 0.024 & 0.036 & 0.003 & 0.016 & -0.018 & 0.008\\
15.0--16.9 & $15.9$ &$(5.166 \, \pm 0.010 \, _{-0.259}^{+0.248}) \times 10^{-2}$ & 0.009 & 0.024 & 0.036 & 0.006 & 0.014 & -0.023 & 0.010\\
16.9--18.9 & $17.8$ &$(3.628 \, \pm 0.007 \, _{-0.187}^{+0.176}) \times 10^{-2}$ & 0.009 & 0.024 & 0.036 & 0.009 & 0.013 & -0.026 & 0.011\\
18.9--21.2 & $20.0$ &$(2.524 \, \pm 0.006 \, _{-0.132}^{+0.123}) \times 10^{-2}$ & 0.008 & 0.024 & 0.036 & 0.012 & 0.012 & -0.028 & 0.011\\
21.2--23.8 & $22.5$ &$(1.706 \, \pm 0.004 \, _{-0.090}^{+0.084}) \times 10^{-2}$ & 0.008 & 0.024 & 0.036 & 0.016 & 0.011 & -0.028 & 0.010\\
23.8--26.7 & $25.2$ &$(1.177 \, \pm 0.003 \, _{-0.062}^{+0.059}) \times 10^{-2}$ & 0.009 & 0.024 & 0.036 & 0.018 & 0.011 & -0.028 & 0.008\\
26.7--30.0 & $28.3$ &$(8.076 \, \pm 0.018 \, _{-0.422}^{+0.408}) \times 10^{-3}$ & 0.009 & 0.024 & 0.036 & 0.021 & 0.011 & -0.028 & 0.006\\
30.0--33.7 & $31.7$ &$(5.433 \, \pm 0.014 \, _{-0.252}^{+0.246}) \times 10^{-3}$ & 0.010 & 0.000 & 0.036 & 0.023 & 0.011 & -0.027 & 0.004\\
33.7--37.8 & $35.6$ &$(3.747 \, \pm 0.011 \, _{-0.173}^{+0.172}) \times 10^{-3}$ & 0.010 & 0.000 & 0.036 & 0.024 & 0.011 & -0.027 & 0.001\\
37.8--42.4 & $39.9$ &$(2.620 \, \pm 0.009 \, _{-0.122}^{+0.121}) \times 10^{-3}$ & 0.011 & 0.000 & 0.036 & 0.024 & 0.011 & -0.027 & -0.001\\
42.4--47.5 & $44.8$ &$(1.821 \, \pm 0.007 \, _{-0.085}^{+0.085}) \times 10^{-3}$ & 0.011 & 0.000 & 0.036 & 0.024 & 0.012 & -0.027 & -0.003\\
47.5--53.3 & $50.3$ &$(1.268 \, \pm 0.005 \, _{-0.060}^{+0.059}) \times 10^{-3}$ & 0.012 & 0.000 & 0.036 & 0.024 & 0.012 & -0.028 & -0.006\\
53.3--59.9 & $56.4$ &$(8.778 \, \pm 0.042 \, _{-0.428}^{+0.402}) \times 10^{-4}$ & 0.012 & 0.000 & 0.036 & 0.022 & 0.012 & -0.030 & -0.007\\
59.9--67.2 & $63.3$ &$(6.122 \, \pm 0.033 \, _{-0.308}^{+0.276}) \times 10^{-4}$ & 0.012 & 0.000 & 0.036 & 0.021 & 0.013 & -0.032 & -0.009\\
67.2--75.4 & $71.0$ &$(4.213 \, \pm 0.026 \, _{-0.220}^{+0.187}) \times 10^{-4}$ & 0.012 & 0.000 & 0.036 & 0.018 & 0.013 & -0.034 & -0.011\\
75.4--84.6 & $79.7$ &$(2.953 \, \pm 0.021 \, _{-0.160}^{+0.129}) \times 10^{-4}$ & 0.012 & 0.000 & 0.036 & 0.016 & 0.014 & -0.037 & -0.012\\
84.6--94.9 & $89.4$ &$(2.027 \, \pm 0.016 \, _{-0.114}^{+0.087}) \times 10^{-4}$ & 0.012 & 0.000 & 0.036 & 0.014 & 0.014 & -0.040 & -0.012\\
94.9--106.4 & $100.4$ &$(1.45 \, \pm 0.01 \, _{-0.08}^{+0.06}) \times 10^{-4}$ & 0.012 & 0.000 & 0.036 & 0.011 & 0.015 & -0.043 & -0.013\\
106.4--119.4 & $112.5$ &$(9.92 \, \pm 0.10 \, _{-0.60}^{+0.42}) \times 10^{-5}$ & 0.012 & 0.000 & 0.036 & 0.008 & 0.015 & -0.045 & -0.014\\
119.4--134.0 & $126.3$ &$(6.93 \, \pm 0.08 \, _{-0.44}^{+0.29}) \times 10^{-5}$ & 0.013 & 0.000 & 0.036 & 0.006 & 0.015 & -0.048 & -0.014\\
134.0--150.4 & $141.7$ &$(4.90 \, \pm 0.06 \, _{-0.32}^{+0.20}) \times 10^{-5}$ & 0.013 & 0.000 & 0.036 & 0.004 & 0.015 & -0.050 & -0.014\\
150.4--168.7 & $158.9$ &$(3.43 \, \pm 0.05 \, _{-0.23}^{+0.14}) \times 10^{-5}$ & 0.014 & 0.000 & 0.036 & 0.002 & 0.015 & -0.051 & -0.014\\
168.7--189.3 & $178.6$ &$(2.44 \, \pm 0.04 \, _{-0.16}^{+0.10}) \times 10^{-5}$ & 0.014 & 0.000 & 0.036 & 0.001 & 0.015 & -0.051 & -0.015\\
189.3--212.4 & $200.1$ &$(1.66 \, \pm 0.03 \, _{-0.11}^{+0.07}) \times 10^{-5}$ & 0.016 & 0.000 & 0.036 & -0.000 & 0.015 & -0.051 & -0.015\\
212.4--238.3 & $224.6$ &$(1.17 \, \pm 0.02 \, _{-0.08}^{+0.05}) \times 10^{-5}$ & 0.016 & 0.000 & 0.036 & -0.001 & 0.015 & -0.049 & -0.015\\
238.3--267.4 & $252.0$ &$(8.04 \, \pm 0.19 \, _{-0.50}^{+0.34}) \times 10^{-6}$ & 0.017 & 0.000 & 0.036 & -0.001 & 0.014 & -0.046 & -0.016\\
267.4--300.0 & $282.6$ &$(5.60 \, \pm 0.15 \, _{-0.33}^{+0.24}) \times 10^{-6}$ & 0.017 & 0.000 & 0.036 & -0.001 & 0.014 & -0.041 & -0.016\\
300.0--336.6 & $317.3$ &$(3.89 \, \pm 0.12 \, _{-0.22}^{+0.17}) \times 10^{-6}$ & 0.018 & 0.000 & 0.036 & -0.001 & 0.013 & -0.034 & -0.017\\
336.6--377.7 & $355.3$ &$(2.71 \, \pm 0.09 \, _{-0.14}^{+0.12}) \times 10^{-6}$ & 0.018 & 0.000 & 0.036 & -0.000 & 0.013 & -0.026 & -0.018\\
377.7--423.8 & $399.2$ &$(1.82 \, \pm 0.07 \, _{-0.09}^{+0.08}) \times 10^{-6}$ & 0.019 & 0.000 & 0.036 & 0.000 & 0.012 & -0.016 & -0.019\\
423.8--475.5 & $447.7$ &$(1.29 \, \pm 0.06 \, _{-0.06}^{+0.05}) \times 10^{-6}$ & 0.019 & 0.000 & 0.036 & 0.001 & 0.012 & -0.004 & -0.021\\
475.5--598.6 & $529.0$ &$(7.46 \, \pm 0.22 \, _{-0.46}^{+0.32}) \times 10^{-7}$ & $_{-0.045}^{+0.014}$ & 0.000 & 0.036 & 0.001 & 0.011 & 0.016 & -0.023\\
598.6--753.6 & $666.2$ &$(3.52 \, \pm 0.14 \, _{-0.28}^{+0.23}) \times 10^{-7}$ & $_{-0.066}^{+0.019}$ & 0.000 & 0.036 & -0.001 & 0.009 & 0.050 & -0.026\\
753.6--948.7 & $841.8$ &$(1.63 \, \pm 0.08 \, _{-0.14}^{+0.16}) \times 10^{-7}$ & $_{-0.071}^{+0.021}$ & 0.000 & 0.036 & -0.007 & 0.008 & 0.090 & -0.029\\
948.7--1194.3 & $1055.3$ &$(6.72 \, \pm 0.50 \, _{-0.54}^{+0.93}) \times 10^{-8}$ & $_{-0.062}^{+0.026}$ & 0.000 & 0.036 & -0.019 & 0.007 & 0.130 & -0.030\\
1194.3--1892.9 & $1454.7$ &$(1.59 \, \pm 0.14 \, _{-0.16}^{+0.30}) \times 10^{-8}$ & $_{-0.075}^{+0.100}$ & 0.000 & 0.036 & -0.047 & 0.006 & 0.152 & -0.024\\
1892.9--3000.0 & $2241.5$ &$(3.41 \, \pm 0.53 \, _{-0.58}^{+0.55}) \times 10^{-9}$ & $_{-0.127}^{+0.038}$ & 0.000 & 0.036 & -0.106 & 0.009 & 0.152 & 0.008\\
3000.0--4754.7 & $3899.0$ &$(4.68 \, _{-1.38}^{+1.97} \, _{-1.39}^{+1.59}) \times 10^{-10}$ & $_{-0.242}^{+0.273}$ & 0.000 & 0.036 & -0.169 & 0.020 & 0.152 & 0.126\\
4754.7--7535.7 & $5641.9$ &$(1.95 \, _{-0.72}^{+1.08} \, _{-0.92}^{+0.74}) \times 10^{-10}$ & $_{-0.454}^{+0.193}$ & 0.000 & 0.036 & -0.127 & 0.036 & 0.152 & 0.286\\
\hline\hline
    \end{tabular}
  \end{center}
\end{table}
\renewcommand{\arraystretch}{1}

\begin{figure}[h!]
\begin{center}
\includegraphics[width=0.7 \linewidth]{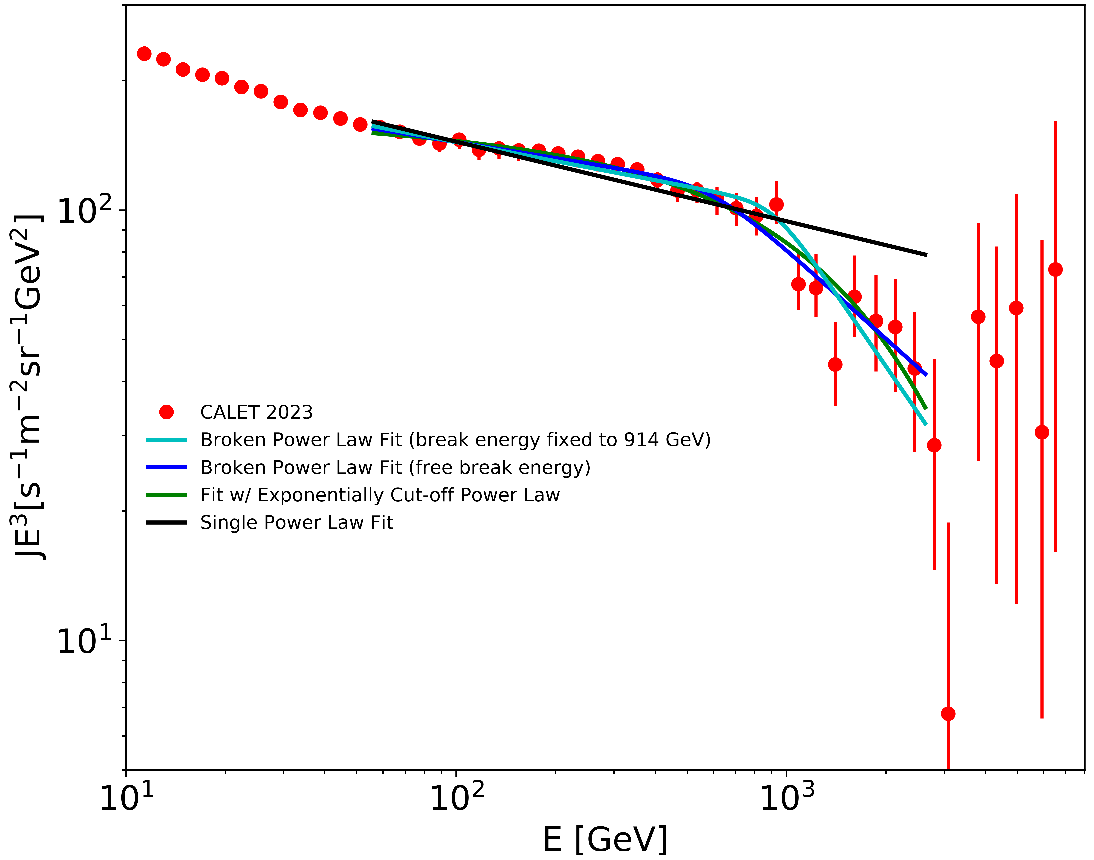}
\caption{All-electron spectrum measured by CALET with the binning used by DAMPE in Ref.~\cite{DAMPE-ae}, and spectral fits  in a restricted energy interval from 55~GeV to 2630~GeV as in our previous publication~\cite{CALET-ae2}  including a broken power law, an exponentially cutoff power law, and a single power law. The error bars represent statistical and systematic uncertainties except normalization. More details can be found in the text. }
\label{fig:FitDampeBin}
\end{center}
\end{figure}

\clearpage
\begin{center}
{\bf FIT OF THE CALET RESULTS WITH DAMPE'S BINNING}
\end{center}
In Fig.~\ref{fig:FitDampeBin}, we show the cosmic-ray all-electron spectrum measured by CALET 
using the same energy binning as DAMPE's result~\cite{DAMPE-ae} and compared with it. The error bars indicate the quadratic sum of statistical and systematic errors  except the normalization uncertainty. 
As in our previous publication~\cite{CALET-ae2}, we fit the spectrum in the energy range from 55~GeV to 2630~GeV with a smoothly broken power law model defined as: \( J(E)=C(E/100 GeV)^\gamma (1+(E/E_b )^{\Delta\gamma/s} )^{-s} \), where $E_b$ is the break energy, while $\gamma $ is the power index below $E_b$ and $\Delta\gamma$ is the difference in the power index below and above $E_b$. The cyan line represents the fit with $E_b$ fixed at 914~GeV as determined by DAMPE ~\cite{DAMPE-ae}, yielding a steepening of the fitted spectrum from $\gamma$=-3.15$\pm$0.01 by $\Delta\gamma$=-0.97$\pm$0.20. The blue line shows a fit in which $E_b$ is a free parameter, yielding a steepening of the fitted spectrum at energy $E_b$ =599$\pm$173~GeV from $\gamma $=-3.12$\pm$0.03 by $\Delta\gamma$=-0.57$\pm$0.18. In both fits, the break smoothness parameter $s$ is fixed at 0.1, and a good fit to our data is obtained, with $\chi^2$/NDF~=~17~/~25 and $\chi^2$/NDF~=~13~/~24, respectively.   An exponentially cutoff power law~\cite{Fermi-ae} (green line) with a power index of -3.03$\pm$0.02 below a cutoff energy of 1921$\pm$243~GeV also fits well, with  $\chi^2$/NDF~=~15~/~25.
In the given energy range and with DAMPE's binning, all these fits are favored at more than 6~$\sigma$ over a single power-law fit, which gives an index -3.19$\pm$0.01 with $\chi^2$/NDF~=~59~/~26.
%

\begin{figure}[h!] 
\begin{center}
\includegraphics[width=0.7 \linewidth]{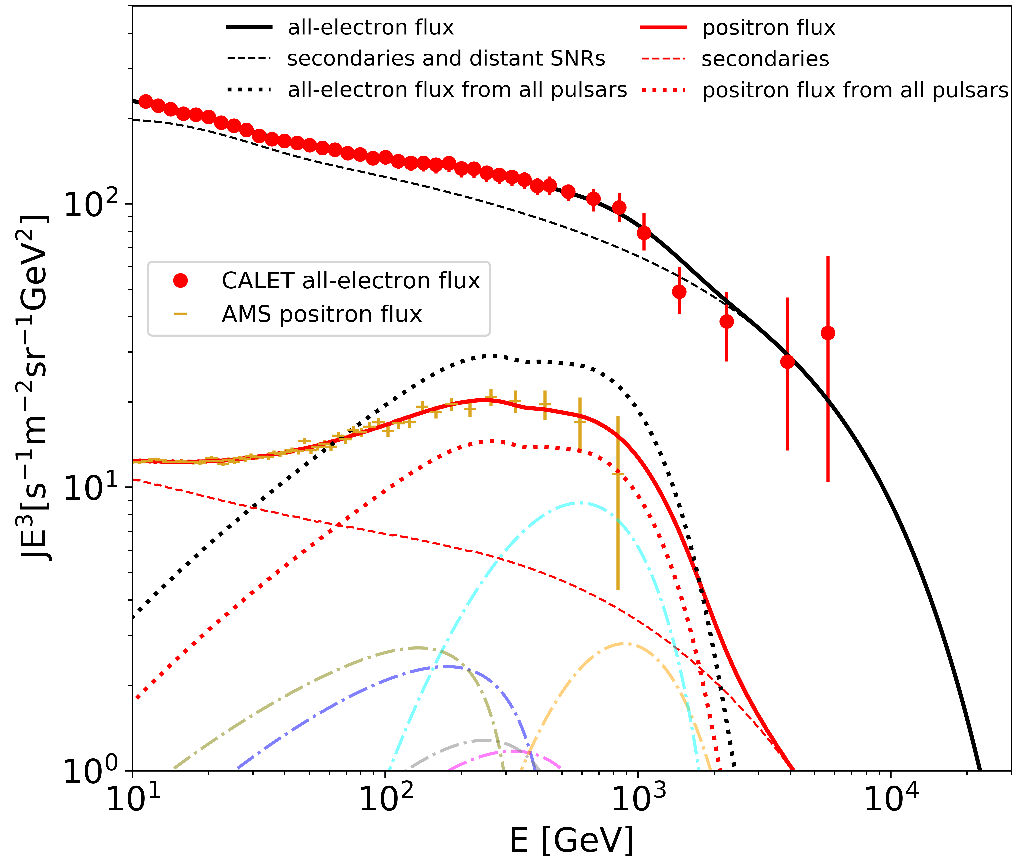}
\caption{Spectral fit over the whole region of the CALET all-electron observation and the AMS-02 positron data as in Fig.4, but without  the contributions from nearby SNRs.}
\label{fig:FitNoSNR}
\end{center}
\end{figure}

\clearpage
\begin{center}
{\bf THE SPECTRUM FITTED WITHOUT CONTRIBUTION FROM NEARBY SNRs}
\end{center}
In Fig.~\ref{fig:FitNoSNR}, we present the best fit over the whole region of the CALET all-electron spectrum and the AMS-02 positron data as in Fig.~4 of the main paper, but without the contribution from the three nearby SNRs. The fits with and without nearby sources give  $\chi^2$/NDF = 34/80 and  $\chi^2$/NDF = 32/80, respectively,  showing that this model fits the data exceedingly well and neither case is significantly  favored over the other. However, the fitting result above 4.8~TeV~(7.5~TeV) predicts an excess of 11.0~(4.2) events with nearby sources and 4.6~(1.0) without nearby sources, respectively.  An excess of 9~(4) events above 4.8~TeV~(7.5~TeV) obtained by the event-by-event analysis described below is compatible with the expected contribution from nearby SNRs (mainly Vela).
\clearpage
\begin{center}
{\bf EVENT-BY-EVENT ANALYSIS: ELECTRON SELECTION WITH LIKELIHOOD ESTIMATION ON EACH EVENT}
\end{center}
For better electron identification above 4.8~TeV, we applied an event-by-event analysis to select the electron candidates which are pre-selected by the BDT response distributions with 13 parameters.  The analysis is carried out for each individual candidate event with a dataset of simulated electrons and protons generated under the same conditions of deposited shower energy,  incident position, and  arrival direction.  The simulated electron dataset for each candidate contains \(10^5\) events thrown with the most likely primary energy to result in the same energy deposit as  the candidate event.  The simulated proton dataset for each candidate contains \(2\times10^5-10^6\) events thrown with a power law distribution (index -2.7) over two orders of magnitude above the real energy deposit.  The analysis procedure follows:
\begin{enumerate}
    \item Filter the simulated datasets using the same pre-selection conditions as applied to the flight dataset in the generation of the candidate list (except for the final BDT selection).
    \item Further filter the simulated proton dataset to only accept events with energy deposit within 50\% of that for the real candidate event.  Energy-deposit dependence of the selection parameters is evaluated (and found to be small) and scaled to the real candidate event's energy deposit.
    \item Generate distributions of the 13 selection parameters for the simulated and filtered electron and proton datasets.
    \item Generate distributions of likelihood ratio (\(LR\)) for the simulated electron and proton datasets according to
      \begin{equation}
        LR(k)=\textrm{log}_{10}\prod_{i=1}^n \frac{p_e^{(k)}(i)}{p_p^{(k)}(i)}  
      \end{equation}
    for event \(k\) and selection parameter \(i\) (up to \(n=13\)), such that
      \begin{equation}
          p_s^{(k)}(i)=\frac{m_s^{(k)}(i)}{m_s^{(tot)}(i)}
      \end{equation}
    where \(m_s^{(k)}(i)\) is the number of events in the same bin as event \(k\) of the histogram of selection parameter \(i\) for species \(s\), and \(m_s^{(tot)}(i)\) is the total number of events in that histogram.
    \item Calculate the \(LR\) for the real candidate event in the same fashion, using the real observed values of the 13 selection parameters.
    \item Scale the distributions of \(LR\) for the electron and proton distributions such that the ratio of protons to electrons matches that observed in the template fit of the pre-cut BDT parameters in the energy bin corresponding to the real candidate event.
    \item Further scale the distributions of \(LR\) such that the sum of simulated electron events with \(LR\) equal to or above that of the real candidate event is equal to 1.  After this scaling, the integral of the proton distribution with \(LR\) equal to or above that of the real candidate event provides the residual proton contamination probability \(p_{cont}\) for that candidate.  That is to say, for each 1 electron in the electron dataset at the confidence level of the real candidate event, \(p_{cont}\) protons are observed.
    \item In order to avoid underestimation of the proton background due to the paucity of proton events surviving at high values of \(LR\), a Gaussian is fit to the tail of the distribution.  The integral of this fit is used as a more conservative estimate of the proton contamination probability.
\end{enumerate}

\begin{figure}[th!]
\begin{center}
\includegraphics[width=0.6 \linewidth]{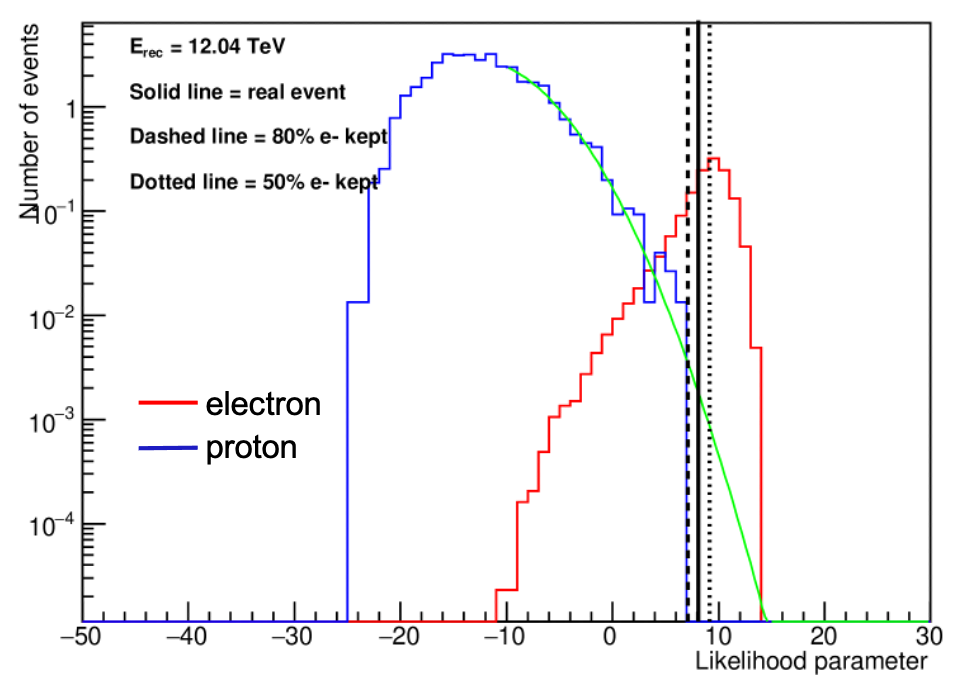}
\caption{Comparison of the LR of the observed event with energy 12.04 TeV with simulated LR distributions of electrons (red) and the corresponding background protons (blue). The green line is a Gaussian fit of the proton distribution tail. The solid vertical line presents the \(LR\) value of the observed event. The dashed line and the dotted line indicate the values of \(LR\) such that 80\%  and 50\%  of the electron sample remain, respectively. }
\label{fig:LR-HE022}
\end{center}
\end{figure}

\begin{figure}[th!]
\begin{center}
\includegraphics[width=0.8 \linewidth]{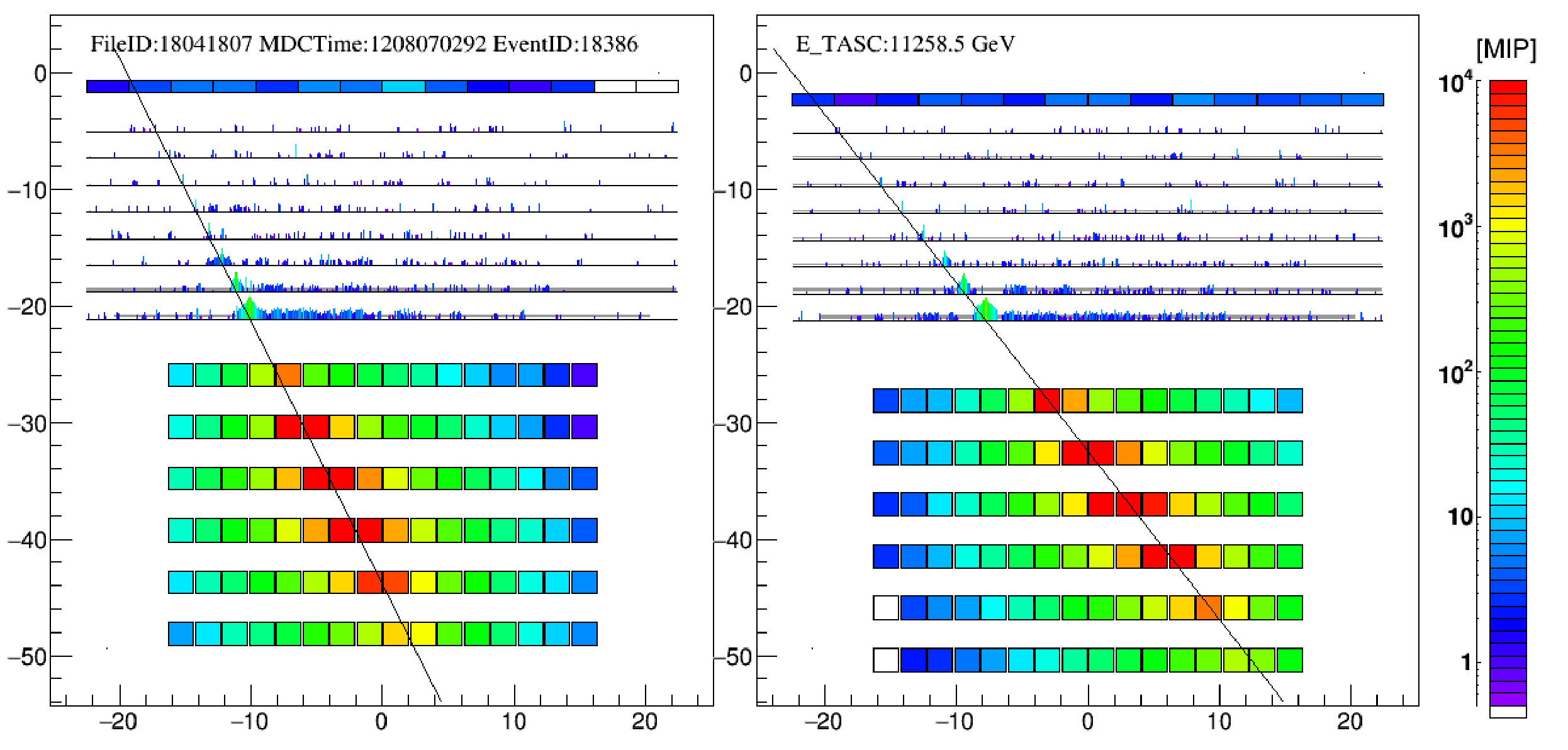}
\caption{The shower profiles projected onto the X-Z (left) and Y-Z (right) view of the  high-energy electron event with an
energy deposit sum in TASC of 11.26 TeV. Black lines represent the
reconstructed tracks using the shower tracking~\cite{emtrack}, which achieves a fine resolution taking advantage of the high granularity
of the IMC. }
\label{fig:Disp-HE022}
\end{center}
\end{figure}

As an example, in Fig.~\ref{fig:LR-HE022}, the \(LR\) distribution of an event with reconstructed energy 12.04 TeV and \(LR\) of 8.47 is compared with the distributions for the corresponding simulated electron and proton samples weighted by the BDT template fit as described above.  Vertical lines indicate the values of \(LR\) such that 80\% (dashed) and 50\% (dotted) of the electron sample remain, and the \(LR\) for the real event (solid).  The observed event has \(p_{cont}\) of 0.004, representing a less than 1\% proton contamination probability.  The shower profile projected to the X-Z and Y-Z planes is presented in Fig.~\ref{fig:Disp-HE022}.

By applying this likelihood analysis to all candidate events in the observation time covered by this work, we select 9~(4) events of the pre-selected 16~(8) events above 4.8~(7.5) TeV which have \(p_{cont} < 0.1\).
%
\clearpage
\providecommand{\noopsort}[1]{}\providecommand{\singleletter}[1]{#1}